\documentclass[aps,article,twocolumn,superscriptaddress,english]{revtex4}

\usepackage{times}
\usepackage{graphicx}
\usepackage{amsmath}
\usepackage{amssymb}
\usepackage[monochrome]{color}

\usepackage{amsfonts}
\usepackage[T1]{fontenc}
\usepackage[latin9]{inputenc}
\usepackage{array}
\usepackage{multirow}
\usepackage{esint}
\usepackage{bm}
\usepackage{bbm}
\usepackage{hyperref}
\usepackage{titlesec}
\usepackage{soul}
\usepackage{mathtools}
\DeclarePairedDelimiter\ket{\lvert}{\rangle}
\DeclarePairedDelimiter\bra{\langle}{\lvert}

\begin{document}

\title{Chiral control of quantum states in non-Hermitian spin-orbit-coupled fermions}




\author{Zejian Ren}
\affiliation{Department of Physics, The Hong Kong University of Science and Technology,\\ Clear Water Bay, Kowloon, Hong Kong, China}

\author{Dong Liu}
\affiliation{Department of Physics, The Hong Kong University of Science and Technology,\\ Clear Water Bay, Kowloon, Hong Kong, China}

\author{Entong Zhao}
\affiliation{Department of Physics, The Hong Kong University of Science and Technology,\\ Clear Water Bay, Kowloon, Hong Kong, China}

\author{Chengdong He}
\affiliation{Department of Physics, The Hong Kong University of Science and Technology,\\ Clear Water Bay, Kowloon, Hong Kong, China}

\author{Ka Kwan Pak}
\affiliation{Department of Physics, The Hong Kong University of Science and Technology,\\ Clear Water Bay, Kowloon, Hong Kong, China}

\author{Jensen Li}
\email{jensenli@ust.hk}
\affiliation{Department of Physics, The Hong Kong University of Science and Technology,\\ Clear Water Bay, Kowloon, Hong Kong, China}
\affiliation{IAS Center for Quantum Technologies, The Hong Kong University of Science and Technology, \\ Clear Water Bay, Kowloon, Hong Kong, China}

\author{Gyu-Boong Jo}
\email{gbjo@ust.hk}
\affiliation{Department of Physics, The Hong Kong University of Science and Technology,\\ Clear Water Bay, Kowloon, Hong Kong, China}
\affiliation{IAS Center for Quantum Technologies, The Hong Kong University of Science and Technology, \\ Clear Water Bay, Kowloon, Hong Kong, China}


%
\maketitle
\newpage

\vspace{10pt}

{\bf Spin-orbit coupling is an essential mechanism underlying quantum phenomena such as the spin Hall effect and topological insulators~\cite{Hasan.2010}. It has been widely studied in well-isolated Hermitian systems, but much less is known about the role dissipation plays in spin-orbit-coupled systems~\cite{Ashida.2020}. Here, we implement dissipative spin-orbit-coupled bands filled with ultracold fermions, and observe parity-time symmetry breaking as a result of the competition between the spin-orbit coupling and dissipation. Tunable dissipation, introduced by state-selective atom loss, enables us to tune the energy gap and close it at the critical dissipation value, the so-called exceptional point~\cite{Miri.2019}. In the vicinity of the critical point, the state evolution exhibits a chiral response, which enables us to tune the spin-orbit coupling and dissipation dynamically, revealing topologically robust chiral spin transfer when the quantum state encircles the exceptional point. This demonstrates that we can explore non-Hermitian topological states with spin-orbit coupling.}

An open quantum system that does not conserve energy is effectively described by a non-Hermitian Hamiltonian~\cite{Ashida.2020} and exhibits various counterintuitive phenomena that cannot exist when the system is Hermitian. One such example is the fundamental understanding of non-Hermitian topological matter that may require subtle classification in contrast to the Hermitian topological system, such as iconic topological insulators~\cite{Hasan.2010}. Although extensive theoretical research~\cite{Xu.2017,Gong.2018,Yao.2018wlm,Ashida.2020} and experimental works~\cite{Helbig.2020,Xiao.2020,Weidemann.2020} have been carried out on the non-Hermitian topological band, how the non-Hermitian topological state can be classified remains elusive. In particular, spin-orbit coupling (SOC), a key mechanism driving the non-trivial band topology, has not been investigated in non-Hermitian quantum systems. Recently, considerable efforts have been made in ultracold atoms to explore synthetic SOCs
~\cite{Lin.2011,Wang.2012,Cheuk.2012,Huang.2016} and associated topological bands~\cite{Wu.2016,Song.2018,Song.2019,Wang.2021}, and therefore ultracold atoms offer the unprecedented possibility of studying the non-Hermitian SOC mechanism, a critical step toward realizing non-Hermitian topological phases~\cite{Ashida.2020}.

In this work, we make an important step in this direction by realizing non-Hermitian spin-orbit-coupled quantum gases and oberving a parity-time ($\mathcal{PT}$) symmetry-breaking transition as a result of the competition between SOC and dissipation. We implement synthetic SOC for ultracold fermions~\cite{Lin.2011} together with non-Hermiticity tunable in time. This controllability allows us to investigate how the energy spectrum of a spin-orbit-coupled system changes with dissipation and explore the $\mathcal{PT}$ symmetry-breaking transition across the exceptional point (EP) and its topological nature~\cite{Miri.2019}. Exploring the parameter regime from SOC-dominated to loss-dominated behavior, we identify an EP where the $\mathcal{PT}$ symmetry-breaking transition occurs in a fully quantum regime. Finally, we experimentally probe the chiral property of the EP by encircling it in a parameter space and showing chiral quantum state transfer due to the breakdown of adiabaticity. Our work sets the stage for the experimental study of many-body states in the complex energy bands across the $\mathcal{PT}$ symmetry-breaking transition. Recently, the feasibility of realizing unprecedented phenomena in non-Hermitian atomic systems was noted, including enhanced fermionic superfluids~\cite{Yamamoto.2019}, the generalized Floquet time crystal with dissipation~\cite{Lazarides.2020} and higher-order topological phases~\cite{Luo.2019}. 

This work complements the non-Hermitian phenomena observed in classical systems including topological energy transfer~\cite{Xu.2016vi6,Doppler.2016,Hassan.2017uiq}, enhanced sensing~\cite{Wiersig.2014}, $\mathcal{PT}$ symmetric lasing~\cite{Liertzer.2011zy}, and  novel topological entities such as exceptional rings~\cite{Zhen.2015} and non-Hermitian topological edge states~\cite{Zhao.2019}. Nevertheless, the role that non-Hermiticity plays in the genuine quantum regime ranging from few-body to many-body systems remains largely unexplored. Although recent works have explored non-Hermitian systems within a quantum framework~\cite{Minganti.2019} and have demonstrated the $\mathcal{PT}$ symmetry-breaking transition in non-Hermitian quantum systems, such as photons~\cite{Xiao.2017,ozturk.2021}, superconducting qubits~\cite{Naghiloo.2019}, single electronic spins~\cite{Wu.2019}, {\color{red}exciton-polaritons~\cite{Gao.2015}}, or atoms~\cite{Li.2019,Takasu.2020}, the dynamic evolution of the quantum state near the EP has been limited to the single particle quantum system~\cite{Liu.2020}. Here, we demonstrate the chiral control of the quantum state with ultracold fermions showcasing the many-particle quantum systems.

\begin{figure}
\centering
\includegraphics[width=0.95\linewidth]{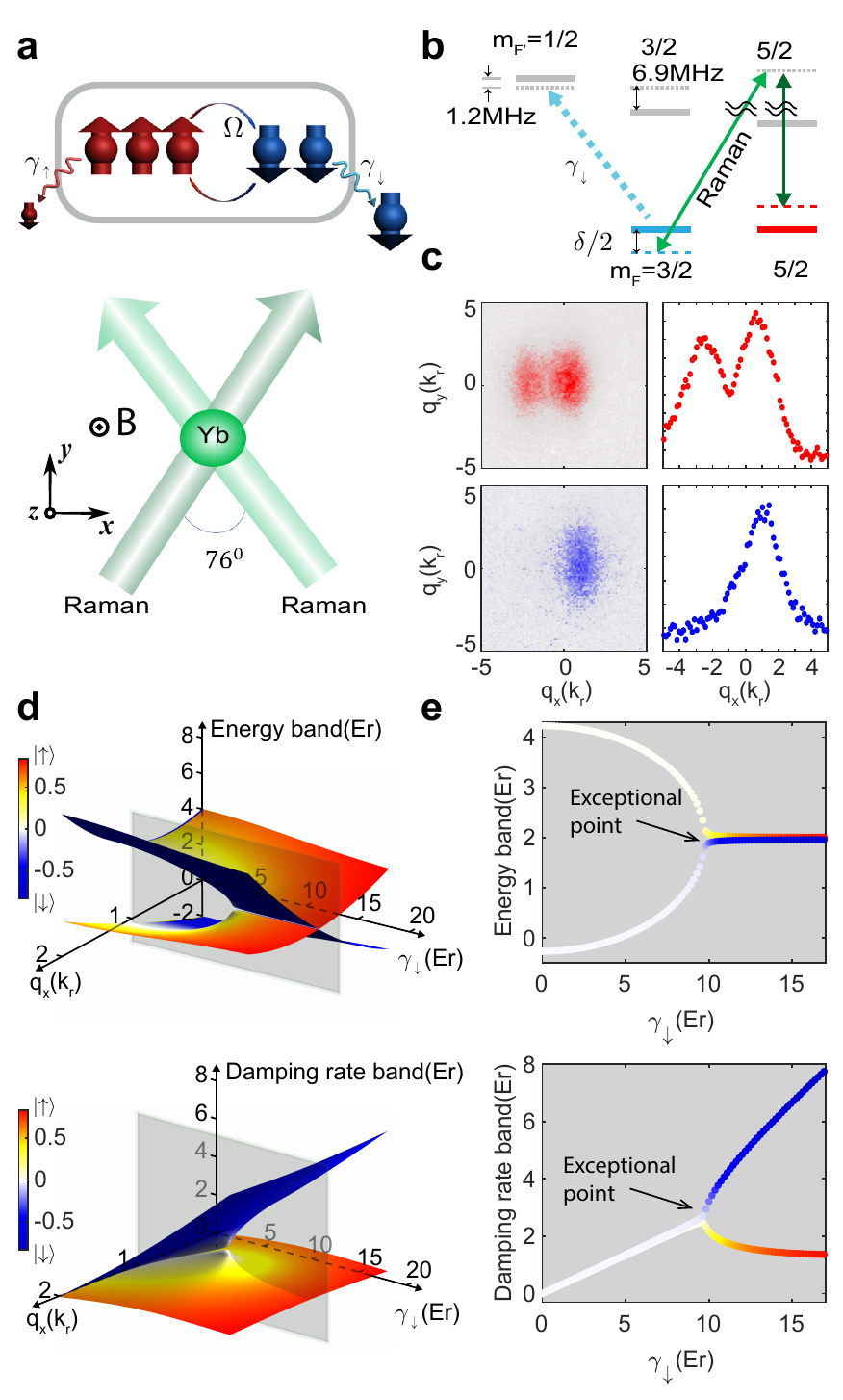}
\caption{ \textbf{Non-Hermitian spin-orbit-coupled system of ultracold atoms }  {\bf a}, Our quantum system comprises ultracold fermions with two spin states, $\ket{\uparrow}$ (red) and $\ket{\downarrow}$ (blue), coupled by Raman beams (green arrow). Atom loss is controlled by additional light (not shown), resulting in spin-dependent loss $\gamma_{\uparrow,\downarrow}$. {\bf b}, Schematic energy level diagram with relevant transitions. The atom loss beam is detuned by 1.2~MHz (6.9~MHz) from the $\ket{m_F=3/2}\to\ket{m_{F'}=1/2}$ transition ($\ket{m_F=5/2}\to\ket{m_{F'}=3/2}$ transition). {\bf c}, Density distribution of $\ket{\uparrow}$ (red) and $\ket{\downarrow}$ (blue) atoms with SOC and its cross-sectional profile along the $x$ direction after 6~ms time-of-flight expansion. {\bf d}, Complex energy bands shown by the real and imaginary parts of $\lambda_{\pm}$ in the parameter space of quasimomentum ($q_x$) and dissipation ($\gamma_{\downarrow}$) for $\delta=4E_r$ and $\Omega_R=4.5E_r$. {\bf e}, Energy band ($\text{Re}(\lambda_{\pm})$) and damping rate (-$\text{Im}(\lambda_{\pm})$) showing the EP at $\gamma_{\downarrow}$=9.75$E_r$ along $q_x=1.0k_r$ (gray plane in (D)) when $2\Omega_R=(\gamma_{\downarrow}-\gamma_{\uparrow})$. }
	\label{fig1}
\end{figure}

\begin{figure*}
	\includegraphics[width=0.85\linewidth]{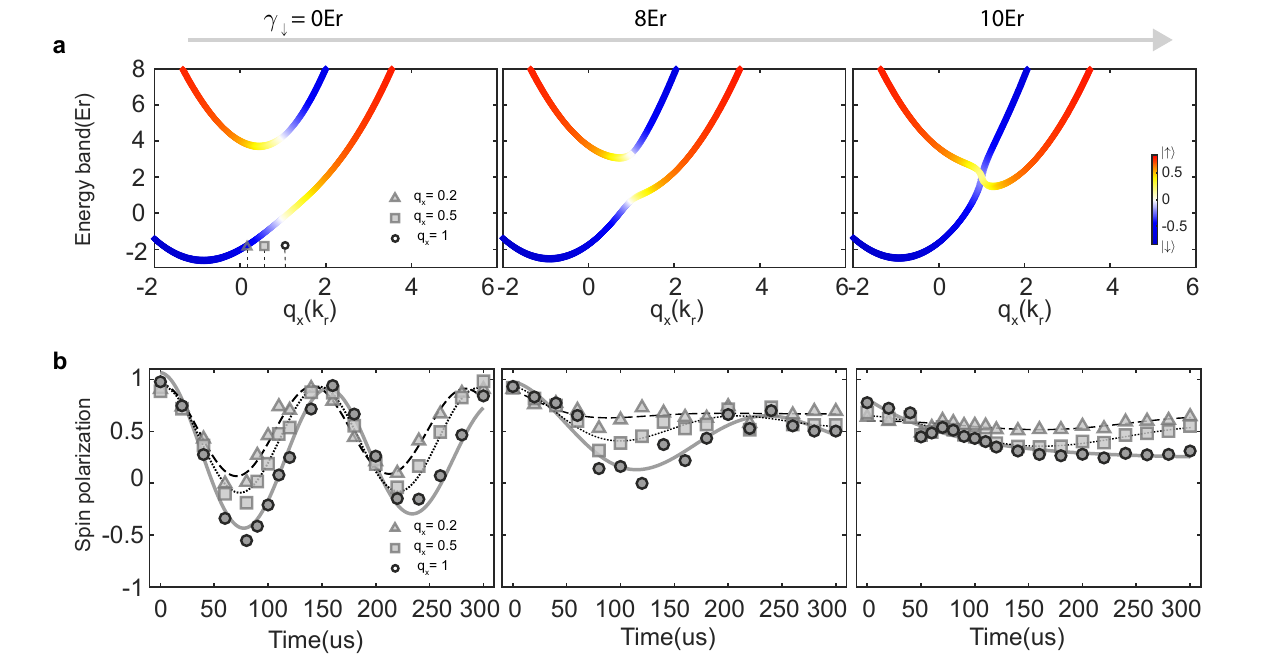}
\caption{ {\bf Momentum-resolved Rabi spectroscopy for observing band merging } {\bf a}, Single-particle energy dispersion of dressed states with increasing dissipation. The energy gap at the resonant quasimomentum $q_x=1.0k_r$ is closed above the EP at $\gamma_{\downarrow}=$9.75$E_r$. {\bf b}, Rabi oscillation for $\delta=$+4$E_r$ and $\Omega_R$=4.5$E_r$. Atoms, prepared in $\ket{\uparrow}$, are suddenly projected into a superposition of eigenstates revealing a nonuniform energy gap. The spin polarization is averaged taking into account the finite optical resolution over $\pm$0.15$k_r$.}\label{fig2}
\end{figure*}

Our experiment begins by loading ultracold fermions of $^{173}$Yb atoms into the engineered energy band~\cite{Song.2018}, in which two hyperfine states (labeled as the $\ket{\uparrow}$ and $\ket{\downarrow}$) are coupled by a pair of Raman transition beams resulting in synthetic SOC with equal strengths of the Rashba and Dresselhaus SOC fields~\cite{Lin.2011}. In a typical energy dispersion, the energy gap between two dressed bands is opened by Raman coupling $\Omega_R$. Adding the spin-dependent atom loss $\gamma_{\uparrow,\downarrow}$, we realize a tunable non-Hermitian Hamiltonian,
$$\mathcal{H}=
\begin{pmatrix}
\frac{\hbar^2}{2m}(q_x- k_r)^2+\delta/2 & \Omega_R/2\\
\Omega_R/2 & \frac{\hbar^2}{2m}(q_x+ k_r)^2-\delta/2
\end{pmatrix}+\mathcal{H}_{loss}$$

where $\mathcal{H}_{loss}=-\frac{i}{2}(\gamma_{\uparrow}\ket{\uparrow}\bra{\uparrow}+\gamma_{\downarrow}\ket{\downarrow}\bra{\downarrow})$, $q_x$ is the quasimomentum of atoms along the $\hat{x}$-direction, $m$ is the mass of the ytterbium atom and $\delta$ is the two-photon detuning. Here, we define natural units of momentum and energy as $\hbar k_r=\sin(\frac{\theta}{2})\frac{2\pi\hbar}{\lambda_{556}}$ and $E_r=\frac{\hbar^2 k_r^2}{2m}$=$h\times$1.41~kHz, where $\lambda_{556}$=556~nm. Then, the real momentum $k_x$ is related to the quasimomentum as $k_x=q_x\mp k_r$ for spin-up and spin-down, respectively. In our experiment, atom loss is induced by the single near-resonant beam (Fig.~\ref{fig1}b), resulting in the fixed ratio of $\gamma_{\downarrow}/\gamma_{\uparrow} = 13$.

The quantum dynamics of two dressed energy bands $\{ \ket{+}, \ket{-}\}$, corresponding to the eigenvalues $\lambda_{\pm}$ of $\mathcal{H}$, are governed by the eigenvalue difference {\color{red}$\Delta\lambda=(\lambda_{+}-\lambda_{-})$.}
In the $q_x$-$\gamma_{\downarrow}$ parameter space , the $\mathcal{PT}$ symmetry-breaking transition occurs at the EP where both eigenvalues and eigenvectors coalesce (Fig.~\ref{fig1}d and e).
When the system is {\it Hermitian} with $\gamma_{\uparrow,\downarrow}$=0, the gap is opened at $q_x=(\frac{\delta}{4E_r})k_r$~\cite{Lin.2011}. With finite atom loss ($\gamma_{\uparrow,\downarrow} \neq$~0), however, the energy gap is reduced and eventually closed at the critical value (i.e., the EP), above which the eigenvalue difference becomes complex as {\color{red}Im($\Delta\lambda$)$\neq$0.} 


We now investigate how the spin-orbit-coupled band is affected by dissipation via momentum-sensitive Rabi spectroscopy. We begin with a spin-polarized degenerate Fermi gas of 2$\times 10^4$ atoms in $\ket{\uparrow}$ without the Raman transition and loss (i.e., $\Omega_R=\gamma_{\uparrow,\downarrow}=0$). Subsequently, $\ket{\uparrow}$ atoms are transferred to the $\ket{\downarrow}$ state when the Raman coupling is switched on (see Fig.~\ref{fig1}c).  Here, brief pulses of the Raman coupling and loss beams are applied for a variable time duration followed by 6~ms ballistic expansion.
The time-of-flight expansion maps momentum to real space, allowing direct momentum resolution of the energy band. 

One of the features associated with dissipation is the closing of the energy gap at the EP~\cite{Zhen.2015}.
Fig.~\ref{fig2}a shows typical spin-orbit-coupled energy bands at different dissipation strengths for $\delta=$4$E_r$ and their
Rabi oscillations pulsing in the Raman field with dissipation for a variable duration (Fig.~\ref{fig2}b), which reveals the nonuniform energy gap of the dressed bands. In the absence of atom loss ($\gamma_{\uparrow,\downarrow}$=0), the state undergoes Rabi oscillation subject to the energy of the gap at each quasimomentum. The Rabi coupling is resonant at $q_x=1.0k_r$, revealing the smallest energy gap. The Rabi oscillation becomes slower with increasing dissipation. By fitting the spin oscillation with a damped sinusoidal function, we determine the energy gap and damping rate at each quasimomentum as described in Fig.~\ref{fig2}b.

\begin{figure*}
	\includegraphics[width=0.95\linewidth]{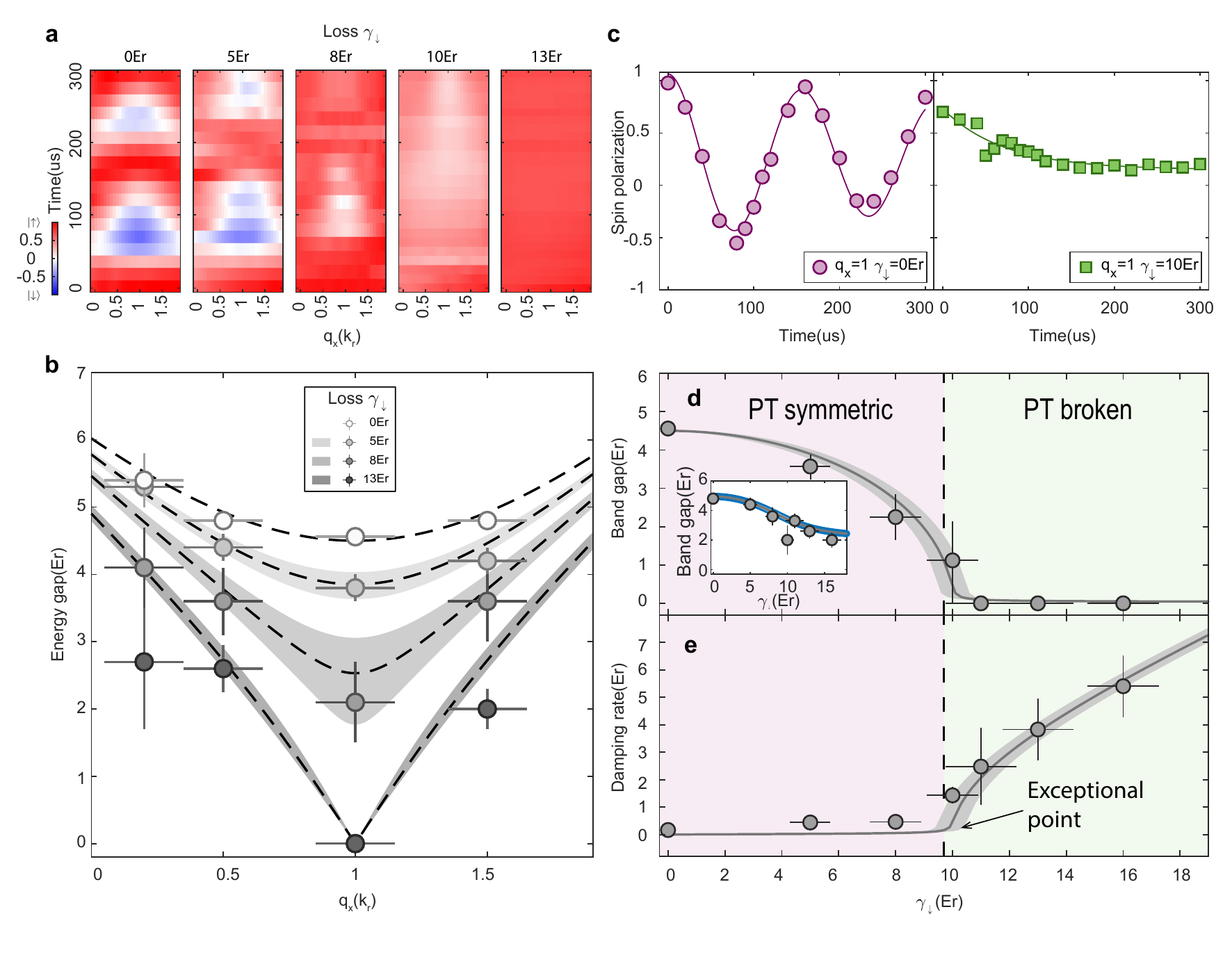}
\caption{\textbf{$\mathcal{PT}$ symmetry-breaking transition and band closing at the EP } {\bf a}, Time-dependent spin texture obtained for different dissipation strengths. The quasimomentum-resolved spin polarization is monitored after switching on SOC fields with dissipation for a variable time. Each spin texture is averaged over 10 measurements. {\bf b}, Energy band gap (circles) measured via Rabi spectroscopy, which is in good agreement with theory (dashed lines). {\color{red}We estimate the uncertainty of the theoretical band gap (shaded region) based on calibrated atom loss and its measurement uncertainty.} {\color{red}{\bf c}, the unbroken phase with prominent Rabi oscillations (left) and the broken phase showing a monotonic spin polarization (right). {\bf d,e} Phase diagram of the $ \mathcal{PT}$ symmetry-breaking transition.} Energy gap and damping rate measured from the Rabi oscillation shown with the real ({\bf d}), and imaginary ({\bf e}), eigenvalues of $\mathcal{H}$ (solid lines), respectively. {\color{red} The shaded region indicates the uncertainty of band gap and damping rate associated with the uncertainty of atom loss.} The error bars in all panels represent standard fitting errors (vertical) and the experimental uncertainty related to the calibration (horizontal).}
	\label{fig3}
\end{figure*}

A Fermi sea collectively undergoes Rabi oscillations in contrast to the non-Hermitian classical~\cite{Miri.2019} or single-particle quantum~\cite{Liu.2020} system, revealing a nonuniform band gap in time-dependent spin textures (see Fig.~\ref{fig3}a).
From this quasimomentum-dependent Rabi oscillation, we extract the energy gap, {\color{red}$\text{Re}(\Delta\lambda)$}, between two dressed bands for different dissipation strengths, as shown in Fig.~\ref{fig3}b, in good agreement with the theoretical expectation. The dissipation gives rise to gap closing at the resonant coupling at $q_x$=1.0$k_r$, while the other quasimomenta still reveal a finite energy gap.

Fig.~\ref{fig3}d and e show a quantitative measurement of the band gap {\color{red}($\text{Re}(\Delta\lambda)$)} and damping rate {\color{red}($\text{Im}(\Delta\lambda)$)} of the non-Hermitian system at $q_x$=1.0$k_r$, manifesting the $\mathcal{PT}$ symmetry-breaking transition. In the $\mathcal{PT}$ symmetric phase, the initial quantum state oscillates between two eigenstates of the non-Hermitian Hamiltonian (see Fig.~\ref{fig3}c). Above the EP, however, the strong dissipation completely closes the energy gap with finite {\color{red}$\text{Im}(\Delta\lambda)\neq$0}, giving rise to monotonic behavior, as shown in Fig.~\ref{fig3}c. Both the band gap and damping rate show the $\mathcal{PT}$ symmetry-breaking transition near the EP at $\gamma_{\downarrow}=$9.75$E_r$ (see Fig.~\ref{fig3}d,e). We can trace the energy gap at a nonresonant momentum, $q_x$=0.5$k_r$, as shown in the inset of Fig.~\ref{fig3}d. In this case, the energy gap saturates to a finite energy gap. {\color{red}With large dissipation, the spin-orbit-coupled band structure becomes similar to free particle dispersion, which may be related to the quantum Zeno effect (see Methods)~\cite{kofman.2000}.}

We now explore topological features near the EP by dynamically encircling it with fermions. Near the EP, the state evolution is direction dependent, resulting in intriguing chiral behavior when encircling the EP due to the breakdown of adiabaticity in non-Hermitian systems~\cite{Miri.2019}. While this topological chirality has been observed in classical systems~\cite{Xu.2016vi6,Doppler.2016,Mailybaev.2005,Uzdin.2011,Wu.2014,Gao.2015,Hassan.2017uiq,Yoon.2018}, such as photonics and acoustics, it remains largely unexplored in quantum systems~\cite{Liu.2020}, especially in many-body quantum systems. In contrast to classical systems, the quantum system with time-varying dissipation may allow for full Hamiltonian engineering capability.

Here, we demonstrate full time-varying control of a non-Hermitian Hamiltonian such that the EP is effectively encircled by ultracold fermions in different directions. This is enabled by the EP occurring at the quasimomentum $q_x=(\frac{\delta}{4E_r})k_r$ where SOC is resonant. This leads to control of the EP position along the $q_x$ axis, effectively with respect to the Fermi sea in the dressed band, by tuning $\delta$. In our experiment, we simultaneously tune the loss ($\gamma_{\downarrow}$) and two-photon detuning ($\delta$) and dynamically explore arbitrary paths in the complex band.

\begin{figure*}
	\includegraphics[width=0.95\linewidth]{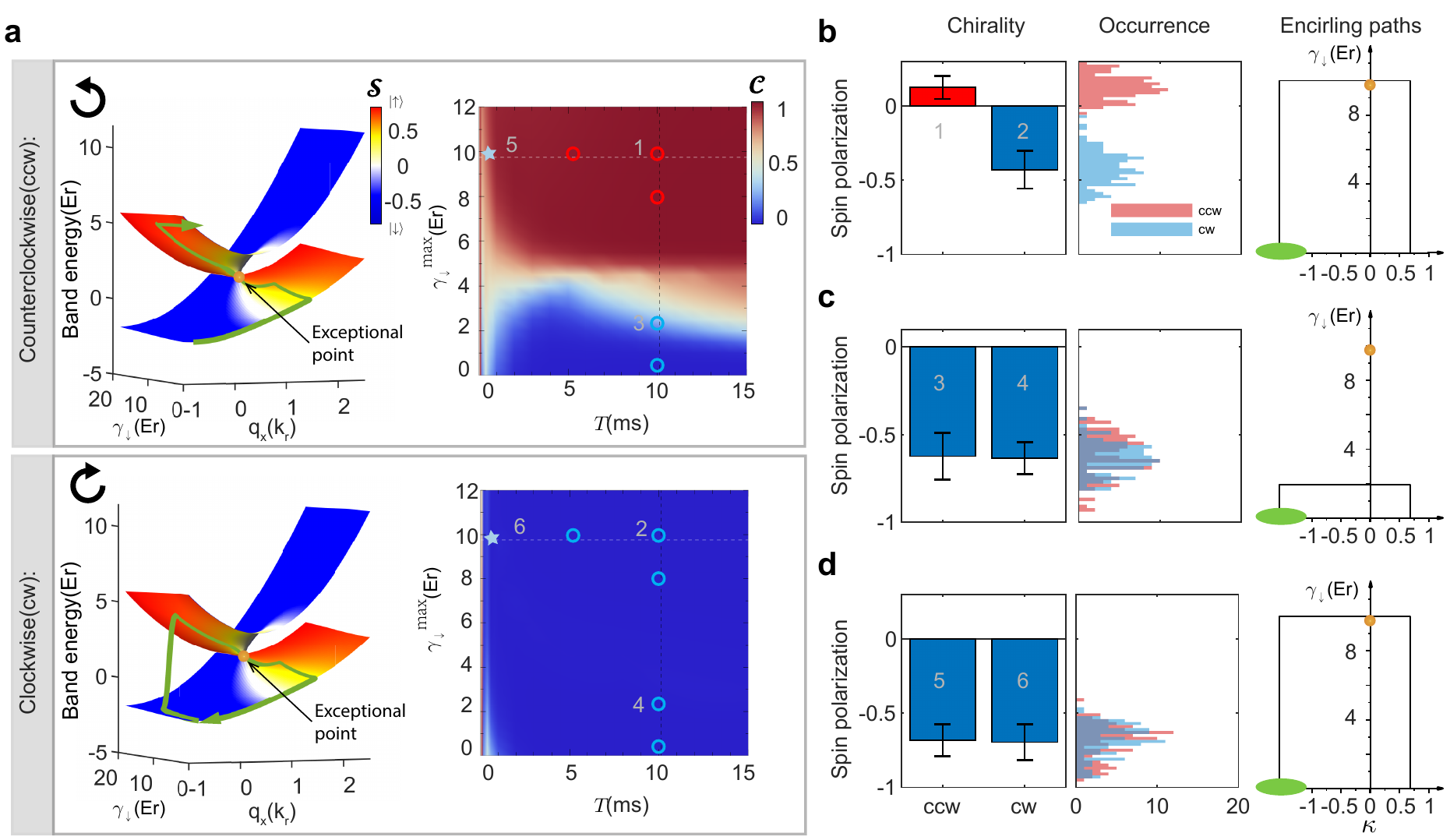}
\caption{\textbf{Topological chiral spin transfer by dynamically encircling an EP } {\bf a}, A Fermi gas is first adiabatically prepared in the initial $\ket{\downarrow}$-dominated state (spin polarization indicated by blue) in the parameter space with $\gamma_{\downarrow}=0E_r$ and $\delta= 3E_r$ (left panel).  We effectively encircle the EP with fermions by tuning the loss and two-photon detuning (green arrow). The spin polarization before dynamic encircling is approximately -0.7 in the initial state. The right panel of (a) shows expected state conversion efficiency $\mathcal{C}$ from $\ket{\downarrow}$ to $\ket{\uparrow}$ against loop duration time T and the maximum atom loss $\gamma_{\downarrow}^{max}$ for the counterclockwise (top) and clockwise (bottom) encircling in the parameter space. The dominant spin polarization of the quantum state measured after the encircling is indicated by the color of the circles and stars. The final spin polarization after encircling is experimentally measured for different $T$ and $\gamma_{\downarrow}^{max}$ as indicated by the colour of the circle or star, after following the trajectory along clockwise (CW) and counterclockwise (CCW) closed loops. {\color{red}The horizontal dashed line indicates the EP.} {\bf b-c}, show the chiral behavior of final spin polarization for the experimental configurations indicated as circles with the same indices in (a).  For other values of $T$, we keep the same fraction of time for each edge of the trajectory as in the experiment. {\bf d}, No chirality is observed when dynamic encircling is instantaneous within 0.04~ms corresponding to solid star in the right panel of (a). For all cases of (b-d), corresponding histograms for the 100 consecutive measurement series with a binning width of 0.02 for spin polarization are shown {\color{red}together with the encircling paths of the Fermi sea (green region) in the parameter space where the exceptional point is indicated by the yellow circle and $\kappa=q_x-\delta/4$ (see the
Supplementary Information for more deatils).} The error bar represents the standard deviation of the measurements. }
	\label{fig4}
\end{figure*}

We begin with adiabatic loading of fermions into the lowest energy band at $\delta=+3E_r$. Initially, a Fermi sea is adiabatically formed at approximately $q_x\simeq -$0.8$k_r$ with the dominant spin component of $\ket{\downarrow}$ (blue color of energy band in Fig.~\ref{fig4}), whereas the EP occurs at $q_x=(\frac{\delta}{4E_r})k_r=$0.75$k_r$ in the $q_x-\gamma_{\downarrow}$ parameter space. We now tune $\delta$ from +3$E_r$ to -6$E_r$, which shifts the EP to $q_x=$-1.5$k_r$. Subsequently, the loss is increased from zero to different $\gamma_{\downarrow}^{max}$, followed by consecutive control of the two-photon detuning and loss. This results in counterclockwise encircling along a closed loop within $T$=10.1~ms in the $q_x-\gamma_{\downarrow}$ parameter space, as depicted in Fig.~\ref{fig4}a (left), where $T$ is the total encircling time (see the Supplementary Materials). By reversing the aforementioned process, we can encircle the EP in a clockwise manner. 

Then, the spin polarization of the quantum state after the encircling is determined via an optical Stern-Gerlach pulse followed by ballistic expansion. When the atom loss $\gamma_{\downarrow}$ is increased to $\gamma_{\downarrow}^{max}$=10 or 8~$E_r$, we observe the initial quantum state $\ket{\downarrow}$ is selectively switched to spin up depending on the encircling direction (Fig.~\ref{fig4}b). This observation reveals the evolution of the quantum state switches the state to a different eigenstate near the EP~\cite{Hassan.2017uiq} depending on the encircling direction. The chiral spin transfer disappears for $\gamma_{\downarrow}^{max}$=2~$E_r$ if the EP is located far from the loop (Fig.~\ref{fig4}c).


To better understand the quantum dynamics during the nonadiabatic evolution, we perform  numerical calculations in a larger domain of accessible parameters in $\gamma_{\downarrow}^{max}$ and encircling duration. {\color{red}The state conversion efficiencies $\mathcal{C}$ for encircling in the two directions are numerically calculated with considering each momentum sector integrated over the Fermi sea, and are plotted as two color maps in Fig.~\ref{fig4}a.} The experimental results in Fig.~\ref{fig4}b and c (labeled 1 to 4) are indicated by circles at $T$=10.1~ms, in good agreement with $\mathcal{C}\gg$~0 for the case 1 and $\mathcal{C}$=0 for the other cases. In fact, a smaller encircling speed (i.e., a larger $T$) increases the chance of nonadiabatic transitions as long as the trajectory runs near the EP and further lowers the transition threshold of $\gamma_{\downarrow}^{max}$~\cite{Hassan.2017uiq}. The chiral behavior is also observed for $T$=5~ms. On the other hand, we find the inability of the system to adapt to the rapidly varying parameters when $T\sim$0.04~ms is small as Fig.~\ref{fig4}d (labeled 5 and 6). {\color{red}The chiral behavior originates from the asymmetry between the (imaginary) dynamical phase factors acquired during the CCW and CW encirclings, which results preferential amplification leading to the imbalanced spin populations.} A more detailed classification of the behavior of $\mathcal{C}$ in different phases is given in the Supplementary Information (Fig.~S4). 

Our system indeed provides an intriguing platform for the study of interplay between many-body statistics and non-Hermiticity. However, the current observations can be effectively described within the non-Hermitian picture by ignoring a quantum jump operator in the Lindblad  equation.  This is an appropriate approximation for a large number of nearly non-interacting atoms, smearing out the effect of the quantum jumps~\cite{Daley.2014,Minganti.2019}. Furthermore, hole excitations of the Fermi sea, which requires a full master equation approach, are rapidly relaxed in the time scale of $E^{-1}_F$ (where $E_F$ is the Fermi energy) due to the decoherence heating process induced by the loss beam. This rapid relaxation allow us to treat the system as the 2$\times$2 non-Hermitian effective Hamiltonian.


In this work, we have experimentally demonstrated how non-Hermiticity affects the dispersion relation of spin-orbit-coupled fermions inducing the $\mathcal{PT}$ symmetry-breaking transition. The topological nature near the EP indicates that non-Hermiticity can fundamentally modify the physical properties of the spin-orbit-coupled quantum system. In contrast to classical systems where only single (bosonic) particle dynamics are considered, our system sets the stage for investigating many-body interacting fermions with dissipation~\cite{Zhou.2020}. Furthermore, the possibility of exploring non-equilibrium dynamics, quantum thermodynamics~\cite{Deffner.2015} and information criticality~\cite{Kawabata.2017} across the $\mathcal{PT}$ symmetry-breaking transition by engineering the non-Hermitian Hamiltonian in a dynamic manner is conceivable.

\noindent {\bf Acknowledgments} G.-B.J. acknowledges support from the RGC, the Croucher Foundation and Hari Harilela foundation through 16305317, 16304918, 16306119, 16302420, C
6005-17G and N-HKUST601/17. J.L. acknowledges support from the RGC through 16304520 and C6013-18G.

\noindent {\bf Competing interests}
The authors declare no competing interests

\noindent{\bf Data  availability}
The data that support the finding of this work are available from the corresponding authors upon reasonable request.\\

\noindent{\bf Author contributions}
Z.R., E.Z., C.H. and K.K.P. carried out the experiment and data analysis and helped with numerical calculations.  D.L. performed theoretical calculations. G.-B.J. and J.L. and supervised
the research.


\newpage
\clearpage

\renewcommand{\thesection}{M-\arabic{section}}
\setcounter{section}{0}  
\renewcommand{\theequation}{M\arabic{equation}}
\setcounter{equation}{0}  
\renewcommand{\thefigure}{M\arabic{figure}}
\setcounter{figure}{0}  

\section*{Methods}
\subsection*{Preparation of the sample } We slow down $^{173}$Yb atoms through the two-stage process and pre-cool them in the intercombination magneto-optical trap. We then begin with the experiment for a two-component degenerate $^{173}$Yb Fermi gas of $2\times 10^4$ atoms and prepared at $T/T_F \lesssim$0.3(1), where $T_F\simeq k_B \times$400~nK, following forced evaporative cooling in an optical dipole trap formed by 1064~nm and 532~nm laser beams with a trap frequency of $\bar{\omega}=(\omega_x \omega_y \omega_z)^{1/3}$ =112$\times$2$\pi$ Hz. {\color{red}Here  $\lvert{\uparrow, \downarrow}\rangle=\lvert{m_F=5/2, m_F=3/2}\rangle$ represent hyperfine states of the ground manifold with a negligible interaction strength of $k_F a_s \simeq $~0.1 where $k_F$ is the Fermi wave vector and $a_s$ denotes the s-wave scattering length.} A quantized axis is fixed by the bias magnetic field of 13.6~G along the z direction. 

To create a non-Hermitian SOC Hamiltonian, two hyperfine states of the ground state manifold of $^{173}$Yb (labeled as the $\ket{\uparrow}$ and $\ket{\downarrow}$ states) are coupled by a pair of Raman transition beams intersecting at $\theta$=76$^{\circ}$ and blue-detuned by $\sim$1~GHz. Typically, before the spin-orbit coupling is switched on by a pair of two-photon Raman beams, two hyperfine levels ($\ket{\uparrow}$,$\ket{\downarrow}$) are isolated from other states (i.e. $m_F=1/2,-1/2,-3/2$ and $-5/2$) using the spin-dependent light shift induced by the $\sigma^-$ polarized beam (referred to a lift beam), blue-detuned by $\sim$1~GHz, which lifts the degeneracy of the ground manifold. {\color{red}Within the experimental resolution, no atoms are observed in the hyperfine states other than $\ket{\uparrow}$ and $\ket{\downarrow}$.} The background heating arising from the Raman transition is negligible within the time scale of the experiment.

\subsection*{Control of atom loss} To control the dissipation of the system, we use a dedicated plane-wave laser beam (referred to loss beam) at a small detuning around the 556~nm narrow intercombination transition ${}^1$S$_0 (F = \frac{5}{2}) \leftrightarrow {}^3$P$_1 (F' = \frac{7}{2})$ with the natural linewidth of 182~kHz. The loss beam is $\sigma^-$ polarized, and is detuned by 1.2~MHz and 6.9~MHz with respect to $\ket{m_F=3/2}\to\ket{m_{F'}=1/2}$ and $\ket{m_F=5/2}\to\ket{m_{F'}=3/2}$ transitions, respectively, which results in spin-sensitive photon scattering. The detuning of the loss beam is chosen such that the light is essentially resonant for atoms in the $\ket{m_F=3/2}=\ket{\downarrow}$ state with the fixed ratio of $\gamma_{\downarrow}/\gamma_{\uparrow}$=13. Although the absorption and reemission of a photon can leave an atom back in its original state, for example, with the probability of $\sim$14$\%$ for atoms in $\ket{\downarrow}$ (see the relative optical transition strength between hyperfine states in Supplementary Information), we define the effective photon scattering rate as the genuine one-body dissipation ignoring the heating effect which is not significant in our current experimental regime. {\color{red}Nevertheless, the sample heats up within the time scale of $E_F^{-1}$ in the strong dissipation regime, for example, at the loss rate of $\gamma_{\downarrow}$=10$E_r$, resulting in $T\simeq$~140~nK.} We expect the heating associated with the photon scattering can be further suppressed by pumping excited atoms in the ${}^3$P$_1$ manifold to ${}^3$S$_1 $ with 680~nm light. We calibrate the photon scattering rate $\gamma_{\sigma}$ by fitting a function $e^{-\gamma_{\sigma} t}$ to the atom number of the state $\ket{\sigma=\uparrow,\downarrow}$ after the loss beam is suddenly switched on (see Supplementary Information for more details). We achieve tunable loss rate by controlling the power of the loss beam. The change in the two-photon detuning ($\delta$) due to the energy shift induced by the loss beam is less than $h\times$0.1~kHz at $\gamma_{\downarrow}=1 E_r$.
\\~\\

\subsection*{$\mathcal{PT}$ symmetric Hamiltonian} {\color{red}The Hamiltonian $\mathcal{H}$ can be decomposed as $\mathcal{H}=\mathcal{H}_{\mathcal{PT}}-i\frac{\gamma_{\downarrow}+\gamma_{\uparrow}}{4} I$ where we define the $\mathcal{PT}$-symmetric Hamiltonian:
\begin{eqnarray*}
  \mathcal{H}_{\mathcal{PT}}&=&\frac{\hbar^2}{2m}(q_x- k_r)^2\ket{\uparrow}\bra{\uparrow}+ 
 \frac{\hbar^2}{2m}(q_x+ k_r)^2\ket{\downarrow}\bra{\downarrow}\\
   & &+(\frac{\delta}{2}+i\frac{\gamma_{\downarrow}-\gamma_{\uparrow}}{4})\sigma_z+\frac{\Omega_R}{2}\sigma_x
 \end{eqnarray*}
Here $\mathcal{P}=\sigma_x$ represents spin exchange operation and $\mathcal{T}=IK$ denotes pseudo time reversal operation where $K$ is the complex conjugate operation which results in $[\mathcal{H}_{\mathcal{PT}}, \mathcal{PT}]=0$ when $q_x=(\frac{\delta}{4E_r})k_r$. The constant decay term $i\frac{\gamma_{\downarrow}+\gamma_{\uparrow}}{4} I$ does not affect the $\mathcal{PT}$-symmetry breaking transition. Alternatively, the gain and loss  can be understood as being balanced at the EP by gauging out $\frac{i}{4}(\gamma_{\uparrow}+\gamma_{\downarrow}){I}$ from $\mathcal{H}$. Near the gap where spin-orbit coupling is resonant (i.e. $q_x=(\frac{\delta}{4E_r})k_r$), the dissipative two-level system is effectively described by $\mathcal{H}_{\mathcal{PT}}$.}

In the strong-dissipation limit (i.e. parity-time symmetry broken phase), the slowly-decaying eigenstate (i.e. complex energy indicated by red color in the manuscript) has a near-unity overlap with the spin-up state without spin-orbit coupling while the rapidly-decaying eigenstate becomes similar to the bare spin-down state. When the dissipation is extremely strong,  the quantum dynamics of slowly-decaying state is effectively projected onto the low-loss manifold, which is a bare energy band of spin-up atoms. Therefore, the spin-orbit coupling effectively disappears in this limit, closing the band gap.

\end{document}